\documentclass[12pt]{article}
\usepackage{graphicx}
\begin{document}

\title{Production of cold fragments in nucleus-nucleus collisions 
in the Fermi-energy domain}

\author{M. Veselsky\\
\\
Institute of Physics, Slovak Academy of Sciences,\\
Dubravska cesta 9, Bratislava, Slovakia\\
e-mail: fyzimarv@savba.sk\\
\\
and\\
\\
G.A. Souliotis\\
\\
Cyclotron Institute, Texas A\&M University, \\ 
College Station, USA
}

\date{}

\maketitle

\begin{abstract}
The reaction mechanism of nucleus-nucleus collisions 
at projectile energies around the Fermi energy is investigated   
with emphasis on the production of fragmentation-like residues. 
The results of simulations 
are compared to experimental mass distributions of elements 
with Z = 21 - 29 observed 
in the reactions $^{86}$Kr+$^{124,112}$Sn at 25 AMeV.  
The model of incomplete fusion is modified and a 
component of excitation energy of the cold fragment 
dependent on isospin asymmetry is 
introduced. The modifications in the model of incomplete fusion 
appear consistent with both overall model framework and available 
experimental data.  
A prediction is provided for the production of very neutron-rich nuclei 
using a secondary beam of $^{132}$Sn where e.g. 
the reaction $^{132}$Sn+$^{238}$U 
at 28 AMeV appears as a possible alternative to the use of fragmentation 
reactions at higher energies. 
\end{abstract}

\section*{Introduction}

Nucleus-nucleus collisions in the Fermi energy domain exhibit 
a large variety of contributing reaction mechanisms and reaction products 
( see e.g. \cite{MVNPA} ) 
and offer the principal possibility to produce mid-heavy 
to heavy neutron-rich nuclei in very peripheral collisions. 
In the reactions of massive heavy ions 
such as $^{124}$Sn+$^{124}$Sn \cite{GSSnSn}  
and $^{86}$Kr+$^{124}$Sn \cite{GSKrSn}, 
an enhancement was observed over the yields expected 
in cold fragmentation which is at present the method of choice to produce 
neutron-rich nuclei. In this case the neutron-rich nuclei are produced in 
damped symmetric nucleus-nucleus collisions with  intense nucleon exchange 
leading to the large width of isotopic distributions. Further enhancement 
of yields of n-rich nuclei was observed 
in the reaction $^{86}$Kr+$^{64}$Ni \cite{GSKrNi} 
in the 
very peripheral collisions, thus pointing to the possible 
importance of neutron and proton density profiles at the projectile and target 
surfaces. In \cite{MVKrNi}, 
the model of deep-inelastic transfer \cite{TG} was supplemented 
with a phenomenological correction introducing the effect of shell structure 
on nuclear periphery. A consistent agreement with experimental data 
was achieved 
in the reactions of a 25 AMeV $^{86}$Kr beam with three different target nuclei, 
specifically allowing to describe the deviation of the nucleon exchange from 
the path toward isospin equilibration.

In this article, we present an investigation of the reaction mechanism 
of the  nucleus-nucleus collisions with an emphasis on the production  
of fragmentation-like residues at projectile energies around the Fermi energy.  
The fragmentation-like residues with mass and charge below 
the projectile can be, according to calculation \cite{MVNPA}, 
produced at projectile energies around the Fermi energy 
in the process of incomplete fusion. The difference to 
high-energy fragmentation as assumed in standard 
abrasion-ablation model (\cite{Gosset,Gaim} ) 
is that here the participant zone fuses 
with one spectator, thus creating a very hot multifragmentation source and the 
other spectator remains cold. The cold projectile-like fragment is not emitted 
at zero angle but due to classical Coulomb trajectory along which the collision 
evolves it appears at larger angles. 
The existence of such a cold fragment can be deduced e.g. from the 
data of Casini et al. \cite{Casini} 
where mass and excitation energy of the projectile-like fragment 
( PLF ) were kinematically 
reconstructed and it was established that the products 
lighter than the projectile are cold and heavier ones 
are hot. Such dependence can not be 
explained using the concept of nucleon exchange as represented by the model 
of deep-inelastic transfer \cite{Rand}.  
The simulation using the incomplete-fusion ( ICF ) code \cite{MVNPA} 
reproduces such behavior reasonably well, 
due to low excitation energy of fragmentation-like residues with masses 
below projectile and considerably higher excitation energy 
of ICF-like residues heavier than projectile. 
Furthermore, a 
large amount of detailed data on hot multifragmentation source in various 
reactions was reproduced, including detailed experimental studies of the 
mid-velocity source \cite{Dore} and single fusion-like source \cite{Frankland}. 
The reaction $^{124}$Sn+$^{27}$Al was used to verify the ICF model using 
heavy residues with masses A=60-90 measured with high precision using 
a fragment separator at forward angles \cite{MVSnAl} 
and both the production cross sections 
and the N/Z trend were reproduced well for these products 
originating from hot projectile-like ICF source. 
However, little is known on N/Z trend of the cold fragments, which can be 
produced in symmetric reactions of heavy nuclei. In the present paper we 
examine such a trend using the experimentally observed heavy residues 
in reactions $^{86}$Kr+$^{112,124}$Sn at projectile energy 25 AMeV and 
refine the model description in order to improve the prediction of the 
production of exotic ( neutron-rich ) nuclei in nucleus-nucleus collisions 
around Fermi energy.

\section*{Production of heavy residues in reactions \\ 
$^{86}$Kr+$^{112,124}$Sn at projectile energy 25 AMeV}

According to the incomplete fusion model \cite{MVNPA}, 
in order to study the production 
of cold fragmentation-like residues around the Fermi energy it is necessary 
to study symmetric nucleus-nucleus collisions of two massive nuclei ( with
target comparable or heavier than projectile ). The products of interest will 
be produced at angles away from zero degrees and the 
measurement at these angles is necessary to observe such products. 
Furthermore, according to the calculation, the cold fragmentation-like residues 
should be increasingly dominant products for the channels with the number 
of stripped protons exceeding seven to eight. Thus the products of primary 
interest are the heavy residues considerably lighter than the 
initial projectile, 
which due to the removal of a significant number of protons 
can be also considerably neutron-rich.    

\begin{figure}[h]
\centering
\vspace{5mm}
\includegraphics[width=11.5cm,height=10.75cm]{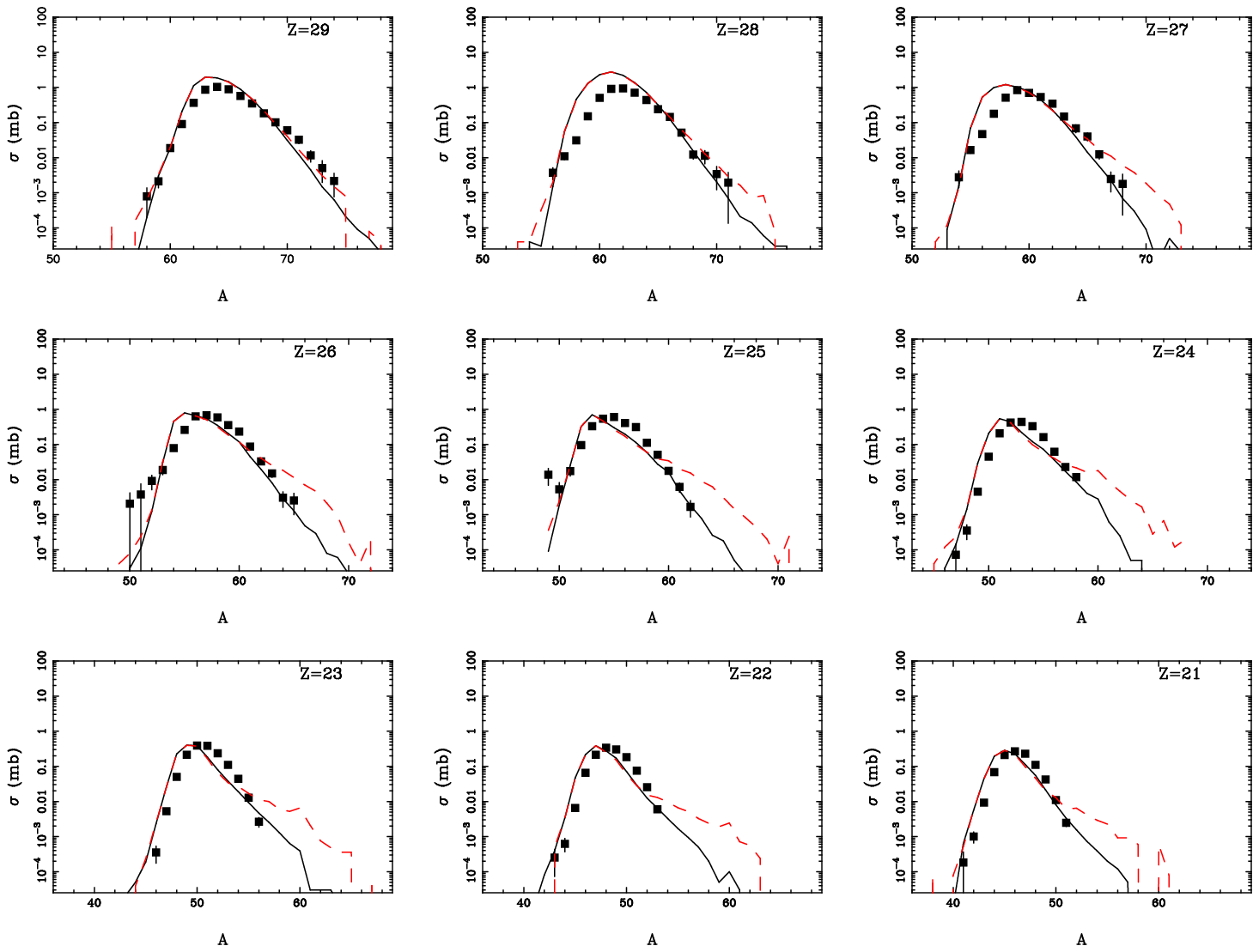}
\caption{\footnotesize
Comparison of the simulations
to experimental mass distributions (symbols ) of elements
with Z = 21 - 29 observed around 4$^{\circ}$ 
in the reaction $^{86}$Kr+$^{124}$Sn at 25 AMeV \cite{GSKrSn}.
Dashed line - results of the standard simulation \cite{MVNPA,MVKrNi}
combined with the de-excitation code SMM \cite{SMM},
Solid line - results of simulation using modified model of incomplete 
fusion ( eq. (\ref{xcaz}) ). 
}
\label{fgkrsn24}
\end{figure}

\begin{figure}[h]
\centering
\vspace{5mm}
\includegraphics[width=11.5cm,height=10.75cm]{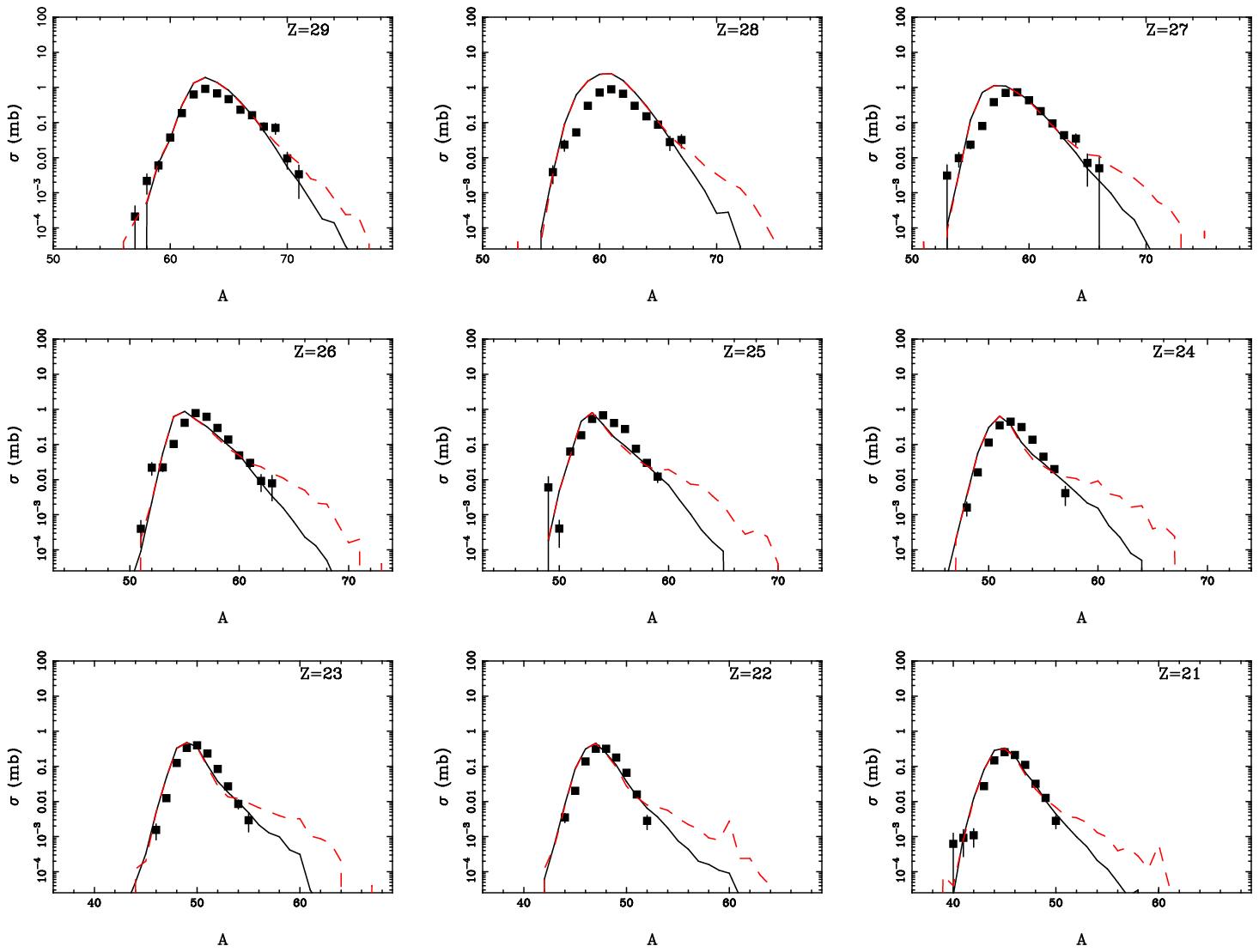}
\caption{\footnotesize
Comparison of the simulations
to experimental mass distributions (symbols ) of elements
with Z = 21 - 29 observed around 4$^{\circ}$ 
in the reaction $^{86}$Kr+$^{112}$Sn at 25 AMeV \cite{GSKrSn}.
Solid, dashed lines - as in Fig. \ref{fgkrsn24}. 
}
\label{fgkrsn12}
\end{figure}

In order to examine the prediction of the ICF code \cite{MVNPA}, 
the results of simulations 
were compared to experimental mass distributions of elements 
with Z = 21 - 29 observed within the separator acceptance 
in the reaction $^{86}$Kr+$^{124,112}$Sn at 25 AMeV \cite{GSKrSn}. 
Figs. \ref{fgkrsn24}, \ref{fgkrsn12} show the comparison of experimental 
mass distributions ( symbols ) to the results of the simulation ( dashed 
line ) for the reactions of $^{86}$Kr with two tin targets. 
The simulation uses either the model of 
deep-inelastic transfer \cite{TG,MVKrNi} 
for peripheral collisions or the model of 
pre-equilibrium emission and incomplete fusion ( ICF ) for violent 
( central ) collisions, combined with the de-excitation code SMM \cite{SMM}.  
The simulated yields were filtered for angular acceptance 
of the separator positioned at $4^{\circ}$ with appropriate azimuthal 
corrections \cite{GSKrSn}. 
The description of experimental data 
using the de-excitation code SMM, representing 
the combination of statistical multifragmentation model for highly 
excited nuclei with evaporation cascade for lower excitation energies, 
proved consistently better than using the model of sequential binary decay. 
Analogous calculation, consisting of the  model of
deep-inelastic transfer for the early stage, combined with the SMM, 
and thus representing only the peripheral collisions, 
was used for the same reactions in the work \cite{MVKrNi} and the yields 
of nuclei with Z = 30 - 35 were reproduced well. The contribution from the 
violent collisions is negligible for nuclei with Z=30-35 and thus 
description of violent collisions used in the present work does not 
affect the results for these reaction products. It is however 
essential for description of the production of nuclei around nickel 
and below where the ICF contribution becomes increasingly dominant.  
One can see that for the reaction $^{86}$Kr+$^{124}$Sn the calculation 
overpredicts the experimental yields of the most neutron-rich products 
below nickel ( with more than 8 stripped protons ). 
The situation is similar also for the reaction $^{86}$Kr+$^{112}$Sn 
even if the most neutron-rich products below nickel are less populated 
than in reaction of $^{86}$Kr+$^{124}$Sn. 
Thus the yields of neutron-rich nuclei appear to be overpredicted 
in the region where the cold fragmentation-like residues dominate. 
A modification of the model of incomplete fusion concerning 
the cold fragmentation-like residues should be considered, 
specifically concerning the description of the N/Z-trend which was not 
verified in the past.   

The model of incomplete fusion \cite{MVNPA}, 
used in the calculation, considers 
a spectator-participant scenario evolving 
along the classical Coulomb trajectory, 
followed by fusion of the participant zone with one spectator, typically the 
heavier one due to larger contact area and thus larger attractive force. 
The charge of spectator zones is determined using the combinatorial 
probability, which is a standard approach in fragmentation 
codes \cite{Gaim}. The excitation energy of the cold fragment is determined 
considering the two-body collisions of participant and spectator nucleons 
along the separation plane \cite{MVNPA}. 
The concept of combinatorial probability explores the available 
statistical phase space and allows a rather wide range of isospin asymmetries. 
From a dynamical point of view, however, the spectator-participant 
scenario implies 
an instantaneous separation and thus preservation of isospin asymmetry 
of the initial nucleus in the ground state, with homogeneous density 
in the interior of the nucleus. Thus, the two concepts seem to be in 
contradiction, which can be resolved when assuming that the change of 
isospin asymmetry is dynamically consistent with transfer of certain 
amount of nucleons across the separation plane. Simultaneously, the 
relative velocity between the participant and spectator zone increases from 
zero to maximum value corresponding to final incomplete-fusion scenario and  
thus the transferred nucleons should carry this relative velocity which 
will be transferred into excitation energy of the acceptor. Thus a 
component of excitation energy dependent on isospin asymmetry can be 
deduced. It can be assumed that, due to absence of Coulomb barrier, 
the nucleons transferred will be predominantly neutrons. 
The formula for such isospin dependent component of excitation energy 
of a cold spectator ( acceptor ) can be written as 

\begin{equation}
\label{xcaz}
E^{*}_S (A_S,Z_S) = x\hbox{ }(A_S - A_{0}(Z_S))\hbox{ }
(\frac{v_{rel}^{ICF}}{v_{proj}})^2 \hbox{ }\frac{E_{P}-V_{C}}{A_{P}}
\end{equation}

where $E_{P}$, $A_{P}$ are the projectile kinetic energy and mass, 
$V_{C}$ is the Coulomb barrier, 
$v_{proj}$, $v_{rel}^{ICF}$ are the projectile velocity and the final relative 
velocity between hot and cold fragment in the incomplete-fusion scenario, 
evaluated at the Coulomb barrier, 
$A_S$, $Z_S$ are mass and charge of the spectator ( cold fragment ), 
$A_0$($Z_S$) is the spectator mass corresponding to N/Z of initial nucleus 
and $x$ is a random number between zero and one, generated for each collision. 
The random number is introduced due to uncertainty concerning the exact 
moment of transfer and represents an zero-th order estimate allowing 
to reproduce the mean value of extra excitation energy due to transfer 
of neutrons. The excitation energy is evaluated only in the case 
when the cold fragment is acceptor ( and thus neutron-rich ). For the loss 
of neutrons no such component is evaluated, thus assuming that neutrons 
close to the Fermi level are lost and no enhancement of intrinsic 
excitation occurs.  

The results of modified calculation employing the formula (\ref{xcaz}) 
are shown in Figs. \ref{fgkrsn24}, \ref{fgkrsn12} as solid lines. 
One can see that the overall agreement with the experimental data is improved 
in the modified calculation compared to the standard one. 
The calculation with modification, which is fully consistent with 
overall description of the nucleus-nucleus collisions and introduces 
no free parameters, describes the trend of the experimental yields 
of the most neutron-rich products below nickel for the reaction 
$^{86}$Kr+$^{124}$Sn, which the initial calculation overpredicts. 
The situation improves also for the reaction $^{86}$Kr+$^{112}$Sn.  
For the most neutron-rich products below chromium 
nuclei occasionally appear less populated than it is predicted 
by the model. This however can be caused 
by missing yield in the experiment due to high background from the initial 
beam, because of increasing 
overlap of the charge states ( in terms of magnetic rigidity ) with charge 
states of the scattered beam. The effect is more pronounced for the reaction 
$^{86}$Kr+$^{112}$Sn due to lower experimental yields. 

\begin{figure}[h]
\centering
\vspace{5mm}
\includegraphics[width=11.5cm,height=7.75cm]{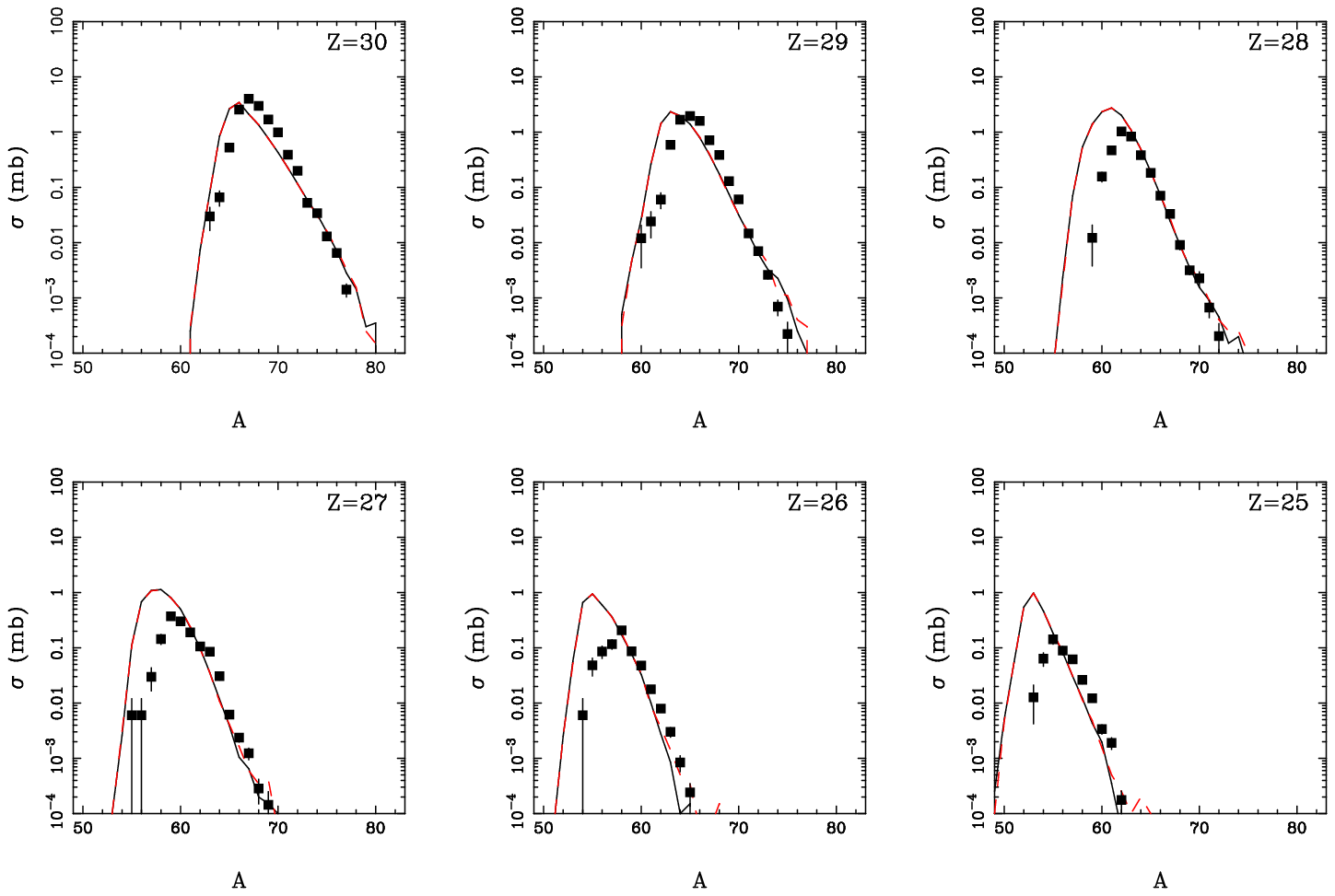}
\caption{\footnotesize
Comparison of the simulated ( lines ) and experimental 
( symbols ) cross sections of heavy residues
with Z = 25 - 30 for the reaction 
of $^{86}$Kr+$^{64}$Ni at 25 A MeV \cite{GSKrNi},
measured at angles below 3 $^{\circ}$. Dashed
and solid line represent standard and modified calculation.
}
\label{fgkrni64}
\end{figure}

It is worthwhile to mention that both the original and modified calculations 
correctly reproduce the experimental cross sections of heavy residues 
with Z = 25 - 30 for the reaction of $^{86}$Kr+$^{64}$Ni 
at 25 A MeV \cite{GSKrNi}, 
measured at angles below 3 $^{\circ}$, as documented in Fig. \ref{fgkrni64} 
( dashed and solid line represent original and modified calculation ). 
The discrepancies at the proton-rich side are caused by restricted 
B$\rho$-coverage for such products in the experiment focused primarily 
on neutron-rich products. The situation in Fig. \ref{fgkrni64} 
and its comparison 
with reactions $^{86}$Kr+$^{112,124}$Sn shows that indeed 
measurement at larger angles and use of heavier target is necessary 
in order to observe the experimental cross section of 
cold fragmentation-like residues. 

In general, the modifications in the model of incomplete fusion 
appear consistent with both overall model framework and experimental 
data and thus one can expect improved predictive power which can be 
used to predict production of exotic mid-heavy to heavy neutron-rich nuclei 
in the reactions around the Fermi energy, and possibly identify under 
which conditions such approach can be more effective than other methods. 
From the point of view of reaction dynamics, the modified model of 
incomplete fusion is consistent with the formation of a neutron-rich 
region between cold and hot fragment ( or participant zone as its 
precursor ). Similar effect was reported in the literature \cite{Neck} as a 
possible consequence of the evolution of nuclear mean field. The number 
of transferred neutrons can then be determined by a mechanism similar to 
the random neck rupture, as established in nuclear fission \cite{RandRupt}, 
which can justify the applicability of a combinatorial ( and thus 
essentially statistical ) probability in the description of dynamical 
reaction mechanism such as the incomplete fusion.   

\section*{Production of neutron-rich nuclei around N=82}

\begin{figure}[h]
\centering
\vspace{5mm}
\includegraphics[width=11.5cm,height=10.75cm]{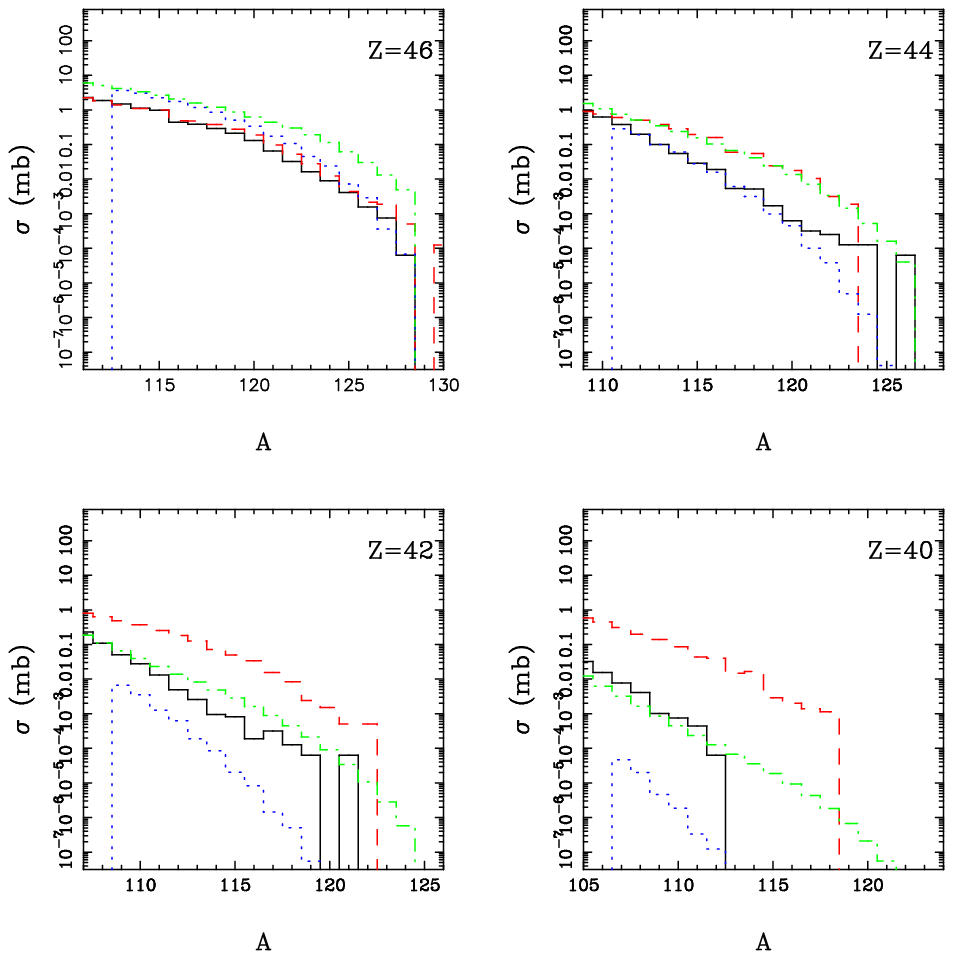}
\caption{\footnotesize
Comparison of production cross sections for reaction
$^{132}$Sn+$^{238}$U at 28 AMeV using standard ( dashed lines )
and modified simulation ( solid lines )
with fragmentation cross sections
of $^{132}$Sn beam with Be target using
COFRA \cite{Cofra} ( dotted lines ) and EPAX \cite{EPAX}
( dash-dotted lines ).
}
\label{fgsn132xs}
\end{figure}

A great deal of attention was paid in recent years to production 
of the secondary beams 
of exotic nuclei. One of the 
most promising ways to produce extremely neutron-rich 
nuclei around the neutron shell N=82 is fragmentation of a secondary beam 
of $^{132}$Sn. Nevertheless, based on the results of the previous section, 
one can 
in principle consider also the reaction in the Fermi-energy domain at energies 
below 50 AMeV. The comparison of production cross sections for the reaction 
$^{132}$Sn+$^{238}$U at 28 AMeV with fragmentation cross section 
of $^{132}$Sn beam with Be target is provided 
in Fig. \ref{fgsn132xs}. 
For the reaction $^{132}$Sn+$^{238}$U the modified DIT code \cite{MVKrNi} 
was used for peripheral collisions together with original 
model of incomplete fusion \cite{MVNPA} ( dashed lines ) 
and its modification presented in this article ( solid lines )
for central collisions, while for the fragmentation of $^{132}$Sn beam 
the codes COFRA \cite{Cofra} ( dotted lines ) and EPAX \cite{EPAX} 
( dash-dotted lines ) were used. 
The production cross sections calculated using both the original and
modified model of incomplete fusion for Z=46 are comparable with results 
of EPAX and COFRA, while for elements with lower 
atomic numbers the reaction $^{132}$Sn+$^{238}$U leads, 
according to both the original 
and modified model of incomplete fusion, to still more favorable 
cross sections exceeding both COFRA and even EPAX cross sections. 
The high cross sections for proton-stripping channels with $^{238}$U target 
were observed recently in reaction $^{58}$Ni+$^{238}$U by Corradi et al. 
\cite{Corr} at energies 
about 1 AMeV above Coulomb barrier. 
The proton-stripping cross sections 
were reproduced using the calculation analogous to the one used here 
after minor readjustment ( increase by 0.5 fm ) 
of diffuseness reflecting mean-field effects 
at such low energy \cite{MVSp2}. When looking at the 
calculated cross sections for $^{132}$Sn+$^{238}$U in Fig. \ref{fgsn132xs} 
one can see that the trend 
of the shapes of mass-distributions calculated using original 
model of incomplete fusion is quite similar to EPAX ( since the 
assumptions are essentially similar ) while trend of the elemental yields 
differs due to assumed classical Coulomb trajectories vs straight movement 
assumed in fragmentation model. The validity of EPAX code concerning the 
trend of n-rich products is a matter of discussion and based on the results 
of the present work one can expect that experimental production cross 
sections of very n-rich nuclei in fragmentation reactions will be 
significantly lower than the prediction of EPAX parametrization. The 
magnitude of such effect needs to be verified ( a dedicated experiment 
on fragmentation of $^{132}$Sn secondary beam is planned 
at GSI \cite{132Prop} ), one can 
nevertheless expect that the code with more realistic treatment of 
spectator excitation energy, such as COFRA, will provide better predictions. 
It is interesting to note that the difference of predictions of fragmentation 
cross sections by EPAX and COFRA is similar to difference of predictions 
of production cross sections in reaction $^{132}$Sn+$^{238}$U 
by the original and
modified model of incomplete fusion. Despite different processes, the 
amount of excitation energy of the spectator appears to be comparable 
in both cases and the success of modified model of incomplete fusion  
in the present work may imply that COFRA will be successful 
in the fragmentation reactions. 

\begin{figure}[h]
\centering
\vspace{5mm}
\includegraphics[width=11.5cm,height=10.75cm]{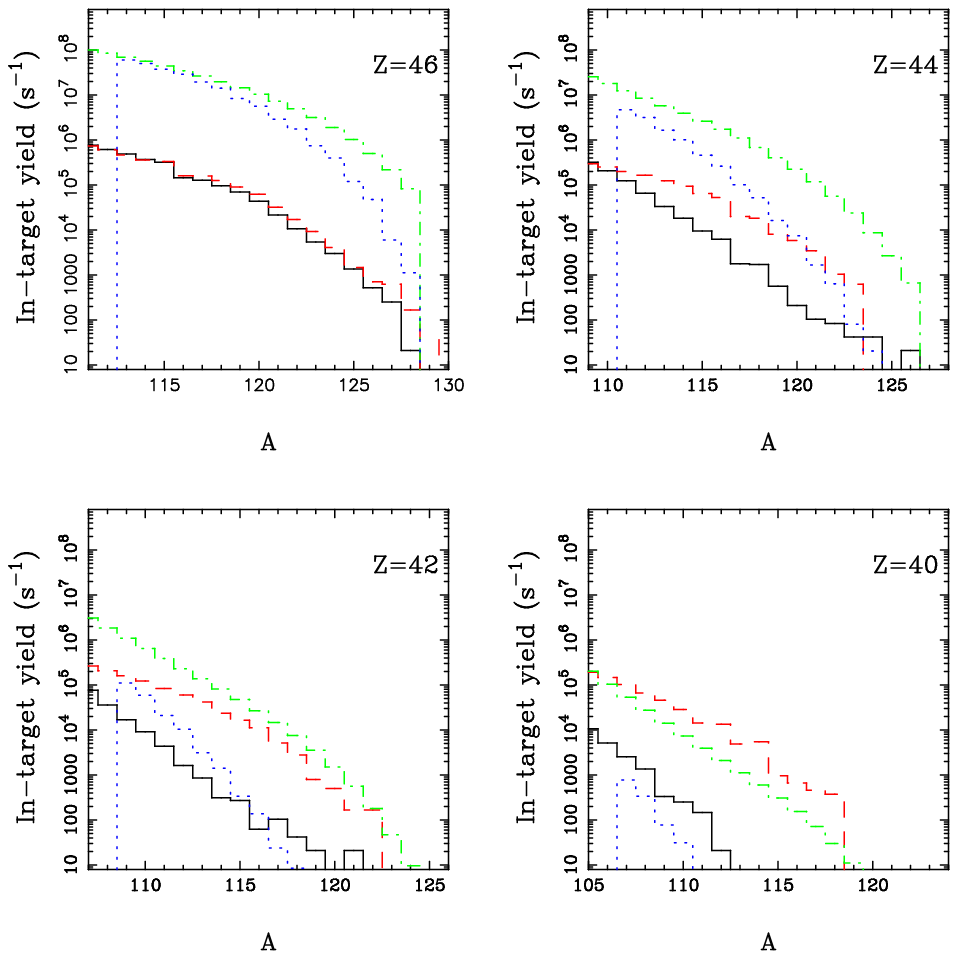}
\caption{\footnotesize
The in-target yields ( for the intensity of $^{132}$Sn 
beam 10$^{12}$ s$^{-1}$ ) calculated using the production 
cross sections from Fig. \ref{fgsn132xs}. Meaning of lines is analogous to  
Fig. \ref{fgsn132xs}.
}
\label{fgsn132tg}
\end{figure}

The in-target yields calculated using the production cross sections 
from Fig. \ref{fgsn132xs} are shown in Fig. \ref{fgsn132tg}. 
For the fragmentation of $^{132}$Sn secondary 
beam an energy 100 AMeV was used, which is the maximum 
foreseen for post-accelerator envisioned for future European RNB facility 
Eurisol \cite{Eurisol}. 
The achievable in-target reaction rate was determined using 
code AMADEUS \cite{Amadeus}. 
For the reaction $^{132}$Sn+$^{238}$U at 28 AMeV  
a target thickness 40 mg/cm$^{2}$ was assumed. 
For the intensity of $^{132}$Sn secondary 
beam a value of 10$^{12}$ s$^{-1}$ was adopted from Eurisol RTD Report 
\cite{Eurisol}. Due to larger target 
thickness, the in-target yield for fragmentation option calculated using 
both COFRA and EPAX dominate for elements Z=44 and above, for lighter nuclei 
nevertheless the larger production cross sections in the Fermi-energy domain 
lead also to larger in-target yields despite relatively thin target and 
for Z=40 the in-target yield calculated using the modified model of incomplete 
fusion exceeds the COFRA value ( and the EPAX value is exceeded by 
original model of incomplete fusion ). To answer on viability of 
Fermi-energy option, a crucial question is which calculation of fragmentation 
cross sections is realistic. Recent measurement carried out at MSU 
\cite{Stolz} has shown 
that experimental fragmentation cross section of neutron-rich Ni isotopes 
are overpredicted by EPAX by up to two orders of magnitude. If the 
situation is analogous in the fragmentation of $^{132}$Sn 
the Fermi-energy option can become 
interesting. However, the angular distribution of reaction products at 28 AMeV 
would require a large-acceptance separator with angular coverage up to 
10 degrees and a highly efficient gas-cell 
in order to form a secondary beam. 

\section*{Conclusions}

The reaction mechanism of the  nucleus-nucleus collisions 
at projectile energies around the Fermi energy was investigated   
with emphasis on the production of fragmentation-like residues. 
The results of simulations 
were compared to experimental mass distributions of elements 
with Z = 21 - 29 observed around 4$^{\circ}$ 
in the reaction $^{86}$Kr+$^{124,112}$Sn at 25 AMeV.   
The model of incomplete fusion was modified and a 
component of excitation energy of the cold fragment 
dependent on isospin asymmetry was 
introduced. The modifications in the model of incomplete fusion 
appear consistent with both overall model framework and available 
experimental data.  
Prediction is provided for the production of very neutron-rich nuclei 
using a secondary beam of $^{132}$Sn where e.g. the reaction 
$^{132}$Sn+$^{238}$U 
at 28 AMeV appears as a possible alternative to the use of fragmentation 
reactions at higher energies. 

The authors acknowledge L. Tassan-Got for providing his DIT code, A.S. Botvina 
for providing his SMM code, K. Summerer for the EPAX-2 routine and 
K.-H. Schmidt and J. Benlliure for providing the results of COFRA 
calculations.  
This work was supported through grant of Slovak Scientific Grant Agency
VEGA-2/5098/25 and by the Department 
of Energy through grant No. DE-FG03-93ER40773.

\end{document}